\documentclass[a4paper]{article}

\usepackage[english]{babel}
\usepackage[utf8x]{inputenc}
\usepackage[T1]{fontenc}

\usepackage[a4paper,top=3cm,bottom=2cm,left=3cm,right=3cm,marginparwidth=1.75cm]{geometry}
\usepackage[backend=biber, maxbibnames=5, giveninits=true, url=false, sorting=none]{biblatex}
\usepackage{amsmath}
\usepackage{amssymb}
\usepackage{graphicx}
\usepackage{xcolor}
\usepackage{url}
\usepackage[colorinlistoftodos]{todonotes}
\usepackage[colorlinks=true, allcolors=blue]{hyperref}
\usepackage{subcaption} 
\usepackage{array}
\usepackage{authblk}

\usepackage[normalem]{ulem} 

\title{Line-of-sight acceleration as a test of the \\Galactic Yukawa potential}

\author[1,2]{Felipe A. da Silva Barbosa}
\author[2]{Luca Amendola}
\author[1,3]{Davi C. Rodrigues}
\author[4]{Roberto Capuzzo-Dolcetta}

\affil[1]{\footnotesize Departamento de Física, Cosmo-Ufes \& PPGCosmo, Universidade Federal do Espírito Santo, 29075-910 Vitória–ES, Brazil}
\affil[2]{\footnotesize Institut für Theoretische Physik, Universität Heidelberg, Philosophenweg 16, 69120 Heidelberg, Germany}
\affil[3]{\footnotesize Centro Brasileiro de Pesquisas Físicas (CBPF), R. Xavier Sigaud 150, 22290-180, Rio de Janeiro, RJ, Brazil}
\affil[4]{\footnotesize Department of Physics, Sapienza, University of Roma, P.le A. Moro 5, 00185 Roma, Italy}

\usepackage{chngcntr} 
\numberwithin{equation}{section} 

\begin{document}

\date{}

\maketitle

\begin{abstract}
We forecast the impact of direct radial acceleration measurements, based on two redshift measurements of the same target one decade apart, on constraining the Yukawa correction to the Newtonian potential in the Milky Way. The Galaxy's matter distribution is modeled as the sum of a spherical bulge, a spherical dark matter halo, and two axially symmetric disks.
Considering a sample of 165 Milky Way globular clusters, we find that the precision of next-generation spectrographs ($\sim$ 10 cm s$^{-1}$) is not sufficient to provide competitive constraints compared to rotation curve data using the same baryonic matter distribution. The latter sample only becomes competitive for a precision better than 0.6 cm s$^{-1}$. On the other hand, we find that adopting a population of $1.3 \times 10^5$ RR Lyrae stars as targets, a precision of  $\sim$ 10 cm s$^{-1}$ can achieve constraints on the Yukawa parameters as strong as with the rotation curves. 
\end{abstract}

\section{Introduction}

Stellar spectroscopy has made considerable advances in the last decades, with the expectation of radial velocity measurements as precise as $10\,  \text{cm} / \text{s}$ in the near future \cite{2020AJ....159..238B}. Many science applications are expected to benefit from such a high precision, among them are the search for exoplanets, the redshift drift in cosmology, the mapping of the Galactic gravitational potential, and, consequently, an inference of the dark-matter distribution \cite{Quercellini:2010zr,Silverwood_Easther_2019, 2018haex.bookE...4W,Chakrabarti_2020, 2023JCAP...11..035F,Ravi_2019,  Chakrabarti_2020}.  Direct measurements of the galactic gravitational potential are possible because  the acceleration induced by the Galactic potential is expected to be $\mathcal{O}(10 \, \text{cm} \, \text{s}^{-1} \, \text{decade}^{-1})$. Such measurements are also preferable to traditional statistical methods, such as Jeans methods, as the former does not rely on assumptions of dynamical equilibrium, although it has its own challenges \cite{Silverwood_Easther_2019,Ravi_2019}. In fact, alternative methods to radial velocity have also been investigated; such methods include eclipsing timing, pulsar timing, and even proposals of using gravitational wave signals \cite{Moran_2024,PhysRevLett.126.141103, Chakrabarti_2022, Ebadi:2024oaq}. Investigations have also been conducted on the possibility of further constraining the gravitational field with angular accelerations \cite{Buschmann_2021}.  In addition to inferring the matter distribution, line-of-sight (LOS) accelerations could also be used to distinguish between different gravity theories and different descriptions of the dark-matter halo, as first suggested in \cite{Quercellini:2008it}, and further developed in \cite{Silverwood_Easther_2019,Chakrabarti_2020, 2023JCAP...11..035F, 2024ApJ...974..223A}.  

In this paper, we attempt to forecast the impact of LOS acceleration measurements in the context of modified gravity. Our focus is on the constraints that can be placed in the Yukawa term as a correction to the Newtonian potential for the baryon-dark matter gravitational interaction.
A Yukawa correction to the Newtonian potential arises generally in scalar-tensor theories within the class of the Horndeski Lagrangian at the linear perturbation level and on sub-horizon scales \cite{DeFelice:2011hq}. In these models, the scalar field mediates an additional interaction between particles.  The same type of correction arises in  bimetric massive gravity \cite{Koennig:2014ods}.
The strength of the Yukawa term depends on the coupling of the scalar field to gravity, while
the interaction range depends on the field's mass. 

Considering the gravitational interactions relevant for baryons and dark matter, baryons and dark matter can have different couplings $\alpha_b$ and $\alpha_{dm}$, respectively. The Yukawa strength $\beta$ of the interaction between species $i$ and species $j$ equals $\alpha_i \alpha_j$. The baryon-baryon gravitational interaction is strongly constrained by local gravity experiments (e.g., \cite{Westphal:2020okx}); the dark matter-dark matter one is of cosmological relevance but is not directly measurable within the Galaxy. The dark matter-baryon interaction is the gravitational interaction due to dark matter that can be probed on galactic scales. Moreover, while the self-interaction $\alpha_i^2$ is bound to be positive for scalar-tensor models\footnote{Although it is negative for vector-mediated interactions.}, the cross-interaction $\alpha_i\alpha_j$ can as well be negative. This paper is devoted to forecasting constraints on this dark matter-baryon interaction. 
As far as we know, this is the first time the LOS Galactic acceleration field is explored in the context of a Yukawa-type correction to the Newtonian potential.
This extra contribution has been studied before in the context of the Galaxy rotation curve \cite{Henrichs_2021}. Here, we extend this work by estimating the expected constraint that comes from LOS accelerations. 

In principle, every star in the Milky Way could be a target to detect the LOS acceleration. Within this huge population, various strategies could be implemented to minimize uncertainties \cite{Silverwood_Easther_2019}. In this paper we select as targets sources that populate the Galaxy halo, namely globular clusters and RR Lyrae stars. 

Our goal is to determine the minimum precision in LOS acceleration measurements such that the resulting constraints on theoretical parameters are comparable to those derived from the rotation curve. This study emphasizes the theoretical principles on distinguishing different gravity theories using the acceleration field, developing further on ref.~\cite{Henrichs_2021}, in particular. However, there are some relevant observational features that we do not address here, in particular deviations from spherical-symmetry of the dark matter profile, uncertainties in the baryonic modeling of the bulge and disk, and substructures in the gravitational potential. Also, before considering true observational data, the perspective acceleration effects need to be taken into account, as explained in the next sections.

\section{Methodology}
\label{methodology}
Our analysis is based on forecasting the sensitivity of LOS acceleration measurements to constrain the parameters of a Yukawa-type correction to the Newtonian potential in the Milky Way. 

We begin by modeling the Galactic potential as the sum of a spherical bulge, two exponential disks (thin and thick), and a spherical dark matter halo described by the Navarro-Frenk-White (NFW) profile. A Yukawa correction is included in the dark-matter-baryon gravitational interaction, characterized by two additional parameters: the coupling strength $\beta$ and the interaction range $\lambda$. 

To determine the fiducial values of the potential parameters, we performed a Bayesian analysis of the Milky Way rotation curve using Markov Chain Monte Carlo (MCMC) sampling with the \texttt{emcee} package \cite{2013PASP..125..306F}. 
The best-fit values serve as the basis for forecasting the constraints achievable from future LOS acceleration data. We then compute the expected LOS acceleration at the position of each target using the gradient of the total gravitational potential projected along the line of sight. The LOS accelerations are simulated for a sample of globular clusters and RR Lyrae stars, using available sets of their spatial distribution. A Gaussian likelihood is adopted to model observational uncertainties, with various assumed precision levels $\sigma_*$ to reflect the capabilities of next generation spectroscopic instruments. For each scenario, we explore the posterior distribution of the potential parameters using MCMC sampling, comparing the resulting constraints with those derived from the rotation curve. This allows us to assess under which observational conditions LOS acceleration data can become competitive or complementary in probing the Galactic potential and modified gravity effects.

\section{Matter distribution and the Yukawa correction in the rotation curve}

\subsection{Baryonic model}\label{sec2.1}

\begin{figure}
        \begin{center}
        \includegraphics[width=0.7 \textwidth]{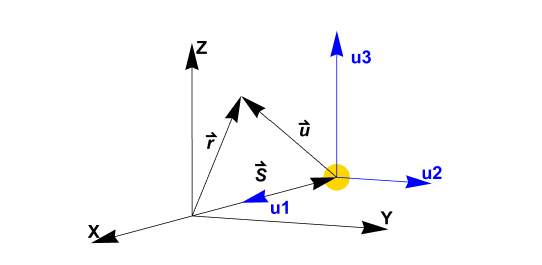}
        \end{center}
        \caption{The adopted coordinate systems. The Galactic rest frame, $(X,Y,Z)$, has its origin on the Galactic center and position vector $\boldsymbol{r}$. The Solar system rest frame, $(\text{u}1, \text{u}2, \text{u}3)$, has its origin in the Sun (yellow dot) and position vector $\boldsymbol{u}$. The Sun's position with respect to the Galactic rest frame is $(X, Y, Z) = (-8.1 \text{ kpc},\,0, \,  0) $.
        }
        \label{coordSystemRelation}
    \end{figure}

We adopt a right-handed coordinate system with the Galaxy center at $(X, Y, Z) = (0,0,0)$, the Solar System at $(X, Y, Z) = (-8.1 \text{ kpc}\,,0,\, 0) $  and the $z$-axis pointing towards the Galactic north pole, the default Galactic frame used in \texttt{astropy}, version version 6.1.3 \cite{astropy:2013, astropy:2018, astropy:2022}.
An illustration of the coordinate system is given in Fig. \ref{coordSystemRelation}. 

Given a Galactic potential, $\Psi(\mathbf{r})$, the acceleration field is
\begin{align}
\ddot{\boldsymbol{r}} = - \nabla \Psi(\boldsymbol{r}) \, ,
\end{align}
where the dots stand for time derivatives.
Since the Solar System rotates around the Galactic center, we define a co-rotating frame with axes aligned to those of the Galactic rest frame and origin at the Sun's position. This setup is also illustrated in Fig. \ref{coordSystemRelation}. 

The rate of change in the LOS velocity is 
\begin{align} \label{eq:udotu}
    \frac{d (\hat{\boldsymbol{u}} \cdot \dot{\boldsymbol{u}})}{dt} = \ddot{\boldsymbol{r}}\cdot \hat{\boldsymbol{u}} - \ddot{\boldsymbol{S}} \cdot \hat{\boldsymbol{u}} +  |\boldsymbol{u}| \, |\dot{\hat{\boldsymbol{u}}}|^2,
\end{align}
where $\boldsymbol{u}$ is the position vector in the Solar system rest frame, $\hat{\boldsymbol{u}}=\boldsymbol{u}/|\boldsymbol{u}|$ and $\boldsymbol{S}$ is the vector from the Galactic center to the Solar system, as in Fig.~\ref{coordSystemRelation}. The last term in Eq.~\eqref{eq:udotu} is the perspective acceleration \cite{1977VA.....21..289V}.
The second term arises from the Solar System's orbital motion around the Galaxy. The first term $\ddot{\boldsymbol{r}} \cdot \hat{\boldsymbol{u}}$  encodes the dynamical effects of the matter distribution and underlying theory of gravity. We define the LOS acceleration $a_{\rm los}$ as
 \begin{align}
    a_{\rm los} \equiv \ddot{\boldsymbol{r}}\cdot \hat{\boldsymbol{u}} =  -\hat{\boldsymbol{u}} \cdot \boldsymbol{\nabla} \Psi_{\text{T}} (\boldsymbol{r}) , 
    \label{los_def}
    \end{align}
    where $\Psi_{\rm T}$ denotes the total potential of the Galactic matter distribution. Thus, our calculations do not account for the Sun's rotation around the Galactic center or the perspective acceleration. However,  real observations should  be corrected for these effects. 

We model the matter distribution as the sum of bulge, thin disk, thick disk, and dark-matter halo components. The Newtonian potentials of the bulge and the two disks are \cite{Eilers_2019},
   \begin{subequations} \label{eq2}
    \begin{align}
    \Psi_{\text{bulge}}(r) &= - \frac{GM }{(r^2 + b^2 )^{1/2}}, \label{eq1a} \\
    \Psi_{\text{disk} }(R,z) &= - \frac{GM }{\left( R^2 + \left[ a + \sqrt{z^2 + b^2 }  \right]^2 \right)^{1/2}  } \, , \label{eq1b}
    \end{align}
    with $r = \sqrt{x^2+y^2+z^2}$  and $R = \sqrt{x^2+y^2}$. The values of the $a$ and $b$ parameters for each of the baryonic components are shown in Table \ref{tableI}.
    \end{subequations}
    \begin{table}
        \centering
        \caption{Mass distribution parameters of the baryonic components. $a$ and $b$ characterize the length scales of the bulge and disk as show in Eq. \eqref{eq2}. Values taken from \cite{Pouliasis_2017} (Model I).}
        \begin{tabular}{ c c c c  } 
        \hline
        \hline
        Component & M [$ 10^9 M_{\odot}$] & $a$ [kpc] & $b$ [kpc] \\ 
        \hline
        bulge & $1.0672$ &  - & $0.3$  \\ 
        thin disk & $3.944$ & $5.3$ & $0.25$\\
        thick disk & $3.944$ & $2.6$ & $0.8$\\
        \hline
        \hline
        \end{tabular}
        \label{tableI}
    \end{table}

\subsection{Dark matter and the Yukawa correction}\label{sec2.2}
We consider a dark matter halo described by the well known NFW profile \cite{Navarro:1996gj, Navarro:2003ew}
    \begin{align}
        \rho (r) = \frac{\rho_s}{\left(\frac{r}{r_s}\right) \left(1 + \frac{r}{r_s} \right)^2} \, ,
    \end{align}
where $r_s$ and $\rho_s$ are constants that can change from galaxy to galaxy.

Following \cite{Henrichs_2021} we consider a Yukawa correction whose presence is associated to a fifth force of scalar nature. For a point particle of mass $M$, the  potential with such correction reads,
\begin{align}\label{eq1}
    \Psi(r) = - \frac{G M}{r} \left( 1+ \beta \, e^{-r/\lambda} \right),
\end{align}
where $\beta$ and $\lambda$ are constants\footnote{Depending on the gravitational model, these constants can be either universal or change from system to system. This distinction is not relevant in our analysis since we focus on a single galaxy.}. 
For a spherically symmetric matter distribution,
\begin{equation} \label{eq:PsiDm}
    \Psi_{\rm dm}(r) = - G \int \frac{\rho(r')}{|\mathbf r - \mathbf r'|} \left( 1 + \beta \, e^{- |\mathbf r - \mathbf r'|/\lambda }\right) \, d^3\boldsymbol{r}' \, .
\end{equation}

Using the NFW profile and the Yukawa correction, the total dark matter potential can be written as \cite{Henrichs_2021}
\begin{subequations}\label{dmDef}
\begin{align}
    &\Psi_{\rm dm}(r) =\Psi_{\rm dm, N}(r) + \Psi_{\rm dm, Y}(r) \, ,\\[.3cm]
    &\Psi_{\rm dm, N}(r) =- \frac{4 \pi  G r_s^3 \rho _s}{r}  \ln \left( \frac{r_s+r}{r_s} \right) \, ,\\[.3cm] \label{dmRCII}
    &\Psi_{\rm dm, Y}(r) = -\frac{2 \pi   G r_s^3 \rho _s \beta}{r } \biggl[e^{\frac{ r_s+r }{\lambda}} \text{Ei}\left(-\frac{r_s + r}{\lambda }\right)-  e^{\frac{ r_s-r }{\lambda }}  \text{Ei}\left(-\frac{r_s}{\lambda }\right) \nonumber\\
    &\qquad \qquad  \qquad \qquad \qquad  - e^{-\frac{r_s+r}{\lambda }}   \text{Ei}\left(\frac{r_s}{\lambda }\right)
    \text{Ei}\left(\frac{r+r_s}{\lambda} \right)  \biggr] \, .
\end{align}
\end{subequations}
In the above, $\Psi_{\rm dm, N}$ refers to the Newtonian part of the dark matter potential $\Psi_{\rm dm}$ (i.e., from Eq.~\eqref{eq:PsiDm} with $\beta =0$) and $\Psi_{\rm dm, Y}$ is the Yukawa correction. The functions Ei are exponential integral functions: ${\rm Ei}(x) = - \int_{-x}^\infty e^{-t}/t \, dt$, where the previous integral is evaluated assuming the Cauchy principal value.

In summary, the full potential is
\begin{align}
    \Psi_T  = \Psi_{\text{bulge}} + \Psi_{\text{thin disk}} +\Psi_{\text{thick disk}} + \Psi_{\text{dm}},
\end{align}
where $\Psi_{\text{bulge}}$ and $\Psi_{\text{disk}}$ are defined in Eq. \ref{eq2} and $\Psi_{\text{dm}}$ is defined in Eq. \ref{dmDef}.
   
    Our strategy is first to derive the best fit of the parameters from real data, i.e. rotation curves, for both the pure Newtonian and the modified gravity models. Then, use these best fits as fiducial to forecast how much can the acceleration field help to distinguish between the two. 
    
\subsection{Rotation Curve}\label{sec3}

Assuming centrifugal equilibrium,  the rotation curve is given by (for $z=0$),
    \begin{align}
        V_c^2(r) =  r \left( \frac{\partial \Psi_{\rm dm}(r)}{\partial r}  +\frac{\partial \Psi_{\text{bulge}}(r)}{\partial r}+\frac{\partial \Psi_{\text{thin disk}}(r,0)}{\partial r} +\frac{\partial \Psi_{\text{thick disk}}(r,0)}{\partial r}  \right).
    \end{align}
    
\begin{figure}
        \begin{center}
        \includegraphics[width=9.179cm, height=6.cm]{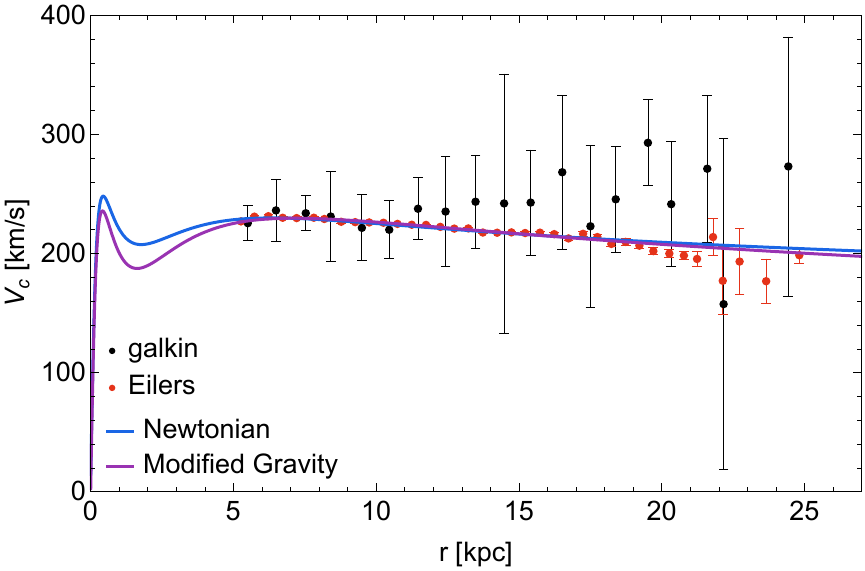}
        \end{center}
        \caption{Best fit and corresponding Galaxy rotation curve of Newtonian, $(\log \rho_s, \log r_s) = (6.95, 1.21)$,  and modified gravity, $(\log\rho_s, \log r_s, \beta, \log \lambda) = (7.28, 1.03, -4, -0.1)$. For this plot the \texttt{galkin} data was binned with bin width of 1 kpc, the error bars are the corresponding standard deviation.}
        \label{RC_BestFit}
    \end{figure}

    For the rotation curve analysis, we use two datasets, Eilers et al. \cite{Eilers_2019} and \texttt{galkin} \cite{Iocco:2015xga, 2017SoftX...6...54P}.  Since the assumption of circular velocity is not expected to be accurate near the galactic center, and given that the velocity errors in the \texttt{galkin} dataset are likely underestimated due to radial uncertainties, as noted in \cite{Henrichs_2021}, we follow \cite{Henrichs_2021} in adopting a minimum rotation velocity uncertainty of 5\%. Additionally, we consider only data points with radius greater than 5 kpc. The data can be visualized in Fig. \ref{RC_BestFit}, the \texttt{galkin} dataset was binned for better visualization, although for sampling we used the full dataset.  
    
    We performed a MCMC sampling, using the \texttt{emcee} package \cite{2013PASP..125..306F}, to obtain the posteriors for the dark matter and modified gravity parameters $(\rho_s, \, r_s, \, \beta, \, \lambda)$. The priors were taken to be uniform in the following ranges:
    \begin{subequations}\label{priors}
        \begin{align}
     5  \times   10^4  M_\odot/\text{kpc}^3< &\rho_s <  5 \times   10^9  M_\odot/\text{kpc}^3,\\
     0.1 \, \text{kpc} < &r_s < 50 \, \text{kpc},\\
    - 15 <  &\beta < 15,\\
     0.1 \, \text{kpc} <  &\lambda < 30 \, \text{kpc}.
    \end{align}
    \end{subequations}
    We find $1 \sigma$ confidence levels consistent with those in \cite{Henrichs_2021}, see  Table \ref{table_unique}. 

\begin{table}
    \centering
    \caption{Inferred parameters from Bayesian analysis. The first column 
    indicates the gravitational theory used in the analysis: Newtonian (Newt.) or modified gravity (M.G.). The second column denotes the observable: Galaxy rotation curve (R.C.), LOS accelerations of globular clusters (G.C.) or LOS accelerations from RR-Lyrae stars (RR-Lyrae). The remaining columns show the inferred parameters results with 1$\sigma$ uncertainty. The central values denote the median, 1$\sigma$ uncertainties are defined from the $0.16$ and $0.84$ quantiles. For the lines using globular clusters as observables, we report results for the smallest adopted $\sigma_a$ (i.e., 0.1 cm s$^{-1}$ decade$^{-1}$) and for the largest value not dominated by the prior.  
    The globular cluster and RR-Lyrae results
    use the best fit in Fig.~\ref{RC_BestFit} as the fiducial model.
    }
    \renewcommand{\arraystretch}{1.5}
\begin{tabular}{ l l  c c  c c c} 
    \hline
    \hline
    Model & Observable &  $\sigma_a$ & $\log_{10} \rho_s $ & $\log_{10} r_s$ & $\beta$ & $\log_{10} \lambda$ \\ 
    &  & \small{$\left[ \frac{\text{cm}}{\text{s} \cdot \text{decade}}\right]$ } & \small{[$M_{\odot} \text{kpc}^{-3}$]} &   [kpc] &  & [kpc] \\
    \hline
    Newt. & R.C. & $- $ & $6.95^{+0.02}_{-0.02}$ &  $1.21^{+0.01}_{-0.01}$ & $-$  & $-$ \\ 
    Newt. & RR-Lyrae & 10. & $6.95^{+0.04}_{-0.04}$& $1.21^{+0.03}_{-0.03}$& $-$ & $-$ \\
    Newt. & G.C. & 0.1 & $6.95^{+0.01}_{-0.01}$ & $1.21^{+0.01}_{-0.01}$ & $-$  & $-$ \\
    Newt. & G.C. & 1.0 & $6.95^{+0.12}_{-0.12}$ & $1.21^{+0.09}_{-0.09}$ & $-$ & $-$ \\[.2cm]
    M.G. & R.C.    & $- $ & $7.28^{ +0.05}_{-0.05}$ & $1.03^{+0.03}_{-0.03}$ & $-7.04^{+3.4}_{-5.61}$ & $-0.23^{+0.14}_{-0.13}$\\
    M.G. & RR-Lyrae & 10. & $7.28^{+0.07}_{-0.07}$ & $1.03^{+0.04}_{-0.04}$ & $-3.98^{+0.33}_{-0.37}$ & $-0.10^{+0.03}_{-0.03}$  \\
    M.G. & G.C. & 0.1 & $7.25^{+0.03}_{-0.04}$ & $1.05^{+0.02}_{-0.02}$ & $-4.20^{+0.20}_{-0.17}$ & $-0.12^{+0.01}_{-0.01}$ \\
    M.G. & G.C. & 0.6 & $7.52^{+0.22}_{-0.21}$ & $1.04^{+0.13}_{-0.12}$ & $-4.09^{+0.96}_{-1.08}$ & $-0.11^{+0.09}_{-0.07}$ \\
    \hline
    \end{tabular}
    \label{table_unique}
\end{table}

\begin{figure}
    \begin{subfigure}[b]{0.35\textwidth}
        \includegraphics[width=\textwidth]{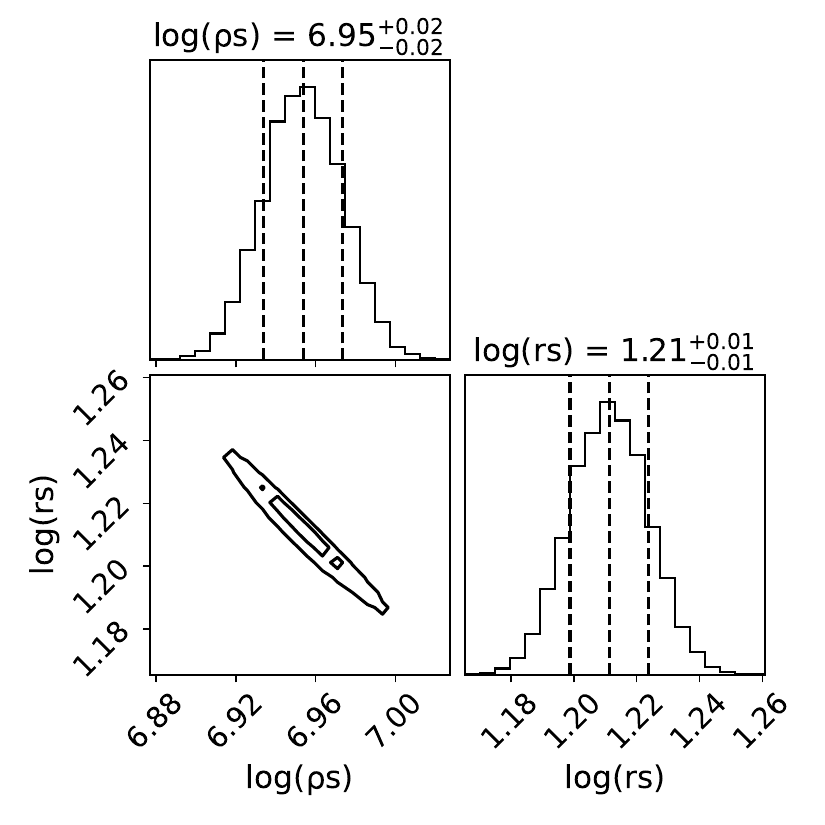}
    \end{subfigure}
    \hspace{1mm} 
    \begin{subfigure}[b]{0.65\textwidth} 
        \includegraphics[width=\textwidth]{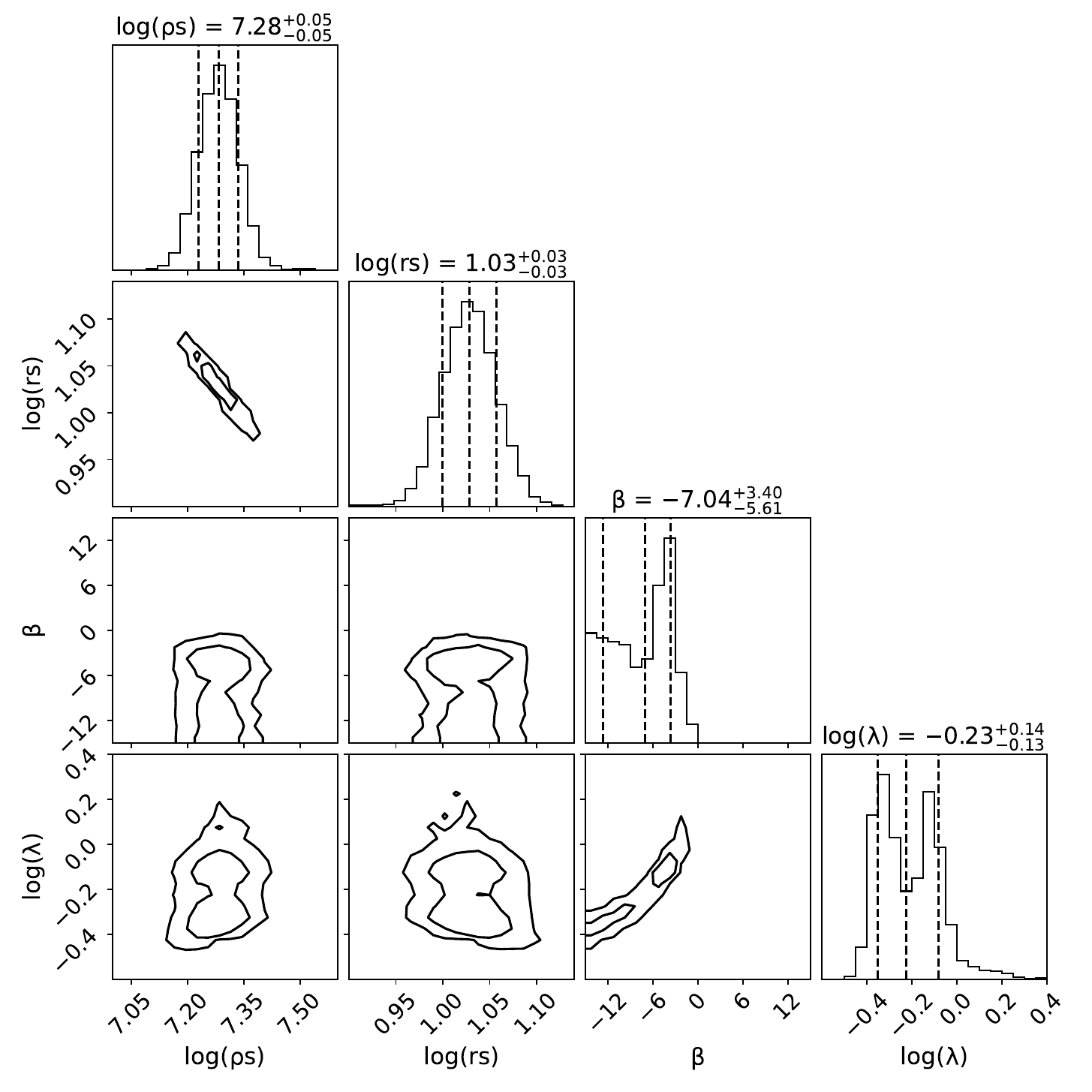}
    \end{subfigure}
    \caption{Rotation curve MCMC analysis. \textit{Left.} The plot assumes Newtonian gravity with NFW dark matter profile with parameters $\rho_s$ [$M_\odot / \text{kpc}^3$] and $r_s$ [kpc]. Vertical dashed lines denote the quantiles 0.16, 0.5, and 0.84, implying a 1$\sigma$ region. \textit{Right.} The plot includes the Yukawa gravity correction, which depends on the parameters $\beta$ and $\lambda$.}
    \label{RCposterior}
\end{figure}

We observe that although a Gaussian approximation is reasonable for the NFW parameters, the parameters associated with modified gravity, $\beta$ and $\lambda$, exhibit strong deviations from Gaussianity behavior, as can be seen in Fig. \ref{RCposterior}. For this reason, we choose to do a full MCMC analysis in the next section instead of recurring to Fisher methods.

\section{Line of sight accelerations}\label{sec4}

\subsection{Statistics}

In order to forecast the constraints from future measurements, we will make use of two kinds of targets as probes of motion in the regime of weak gravity in the Milky Way: globular clusters and RR Lyrae stars. Although globular clusters are not the most abundant targets in the Galaxy, their collective nature could be used to obtain precise measurements of the cluster's center-of-mass velocity, assuming high precision spectrography becomes available for the stars in the cluster. On the other hand, RR Lyrae stars are abundant in the Milky Way: the third Gaia data release \cite{2023ApJ...944...88L} reported more 
than 1.3 $\times 10^5$ RR-Lyrae stars between 0.2 and 138 kpc from the Galactic center, and they could be used to obtain a more detailed map of the Galactic acceleration field.

The adopted Log Likelihood for LOS acceleration is Gaussian, as follows,
\begin{align}\label{Likelihood}
    &\mathcal{L} = - \frac{1}{2} \sum_j \frac{(a_j(\boldsymbol{\theta}_{f}) - a_j(\boldsymbol{\theta}) )^2}{\sigma_{a}^2},\\
    &a_j(\boldsymbol{\theta}) =  -  \hat{\boldsymbol{u}}_j  \cdot \boldsymbol{\nabla}  \Psi_{\text{T}}(\rho_s,\, r_s ,\, \beta,\, \lambda, \boldsymbol{r}_j ) ,\nonumber 
\end{align}
where $a^{\rm (f)}_j$ is the fiducial acceleration of the $j$-th target, the LOS precision  is given by $\sigma_{a}$, $j$ runs over the positions of the targets, $a_{j}(\boldsymbol{\theta})$ is given in Eq. \ref{los_def} and $\boldsymbol{\theta}_f$ is the fiducial point, which we choose to be the best fit of the rotation curve analysis 
\begin{align}
    &\text{Newtonian gravity:} \; \boldsymbol{\theta}_f  = (\log(\rho_s)_f,\, \log(r_s)_f ) =  (6.95, 1.21),\nonumber\\
    &\text{Yukawa gravity:} \;  \boldsymbol{\theta}_f = (\log(\rho_s)_f,\, \log(r_s)_f ,\, \beta_f,\, \log(\lambda)_f) = (7.28, 1.03, -4, -0.1).
\end{align} 
In the next sections we shall present our results for globular clusters and RR Lyrae stars. 

\subsection{Globular clusters}

The globular cluster data were taken from \cite{2021MNRAS.505.5957B}. As previously mentioned the dark matter and Yukawa gravity parameters for the fiducial point were fixed to the best fit value of the rotation curve analysis, Fig. \ref{RC_BestFit}. We used flat priors, as in the rotation curve Eq. \ref{priors}. There are a significant number of globular clusters with total accelerations exceeding $10~\text{cm}~\text{s}^{-1}~\text{decade}^{-1}$: 60 targets based on the accelerations calculated using the Newtonian best fit, and 55 based on those calculated using the Yukawa best fit.

The Newtonian values are typically higher than Yukawa values, which can be understood from the fact that a negative $\beta$ will decrease the acceleration. In particular, we observe that the globular cluster with the highest LOS acceleration  is Djor 2 which is very close to the Galactic center, with $r = 0.8$ kpc.

\begin{figure}
        \begin{center}
        \includegraphics[width=0.49\textwidth]{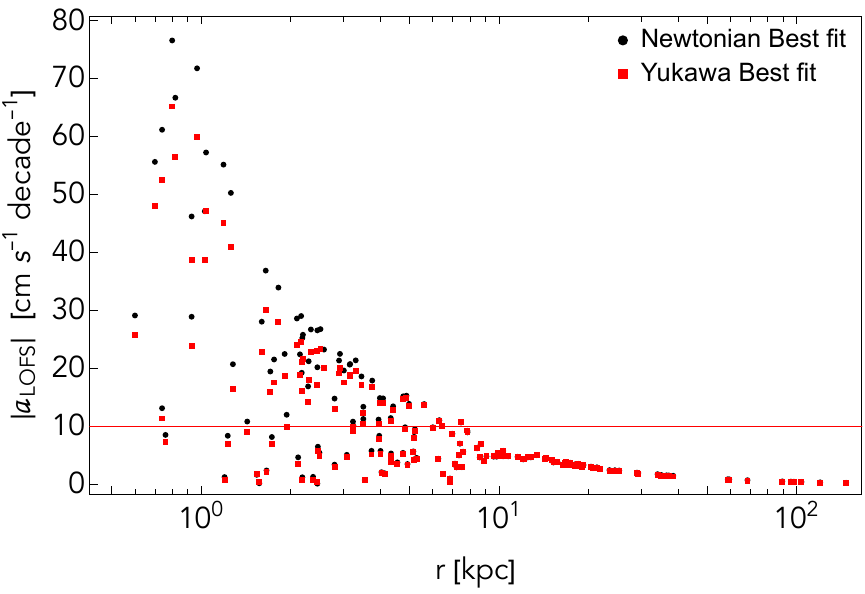}
        \includegraphics[width=0.49\textwidth]{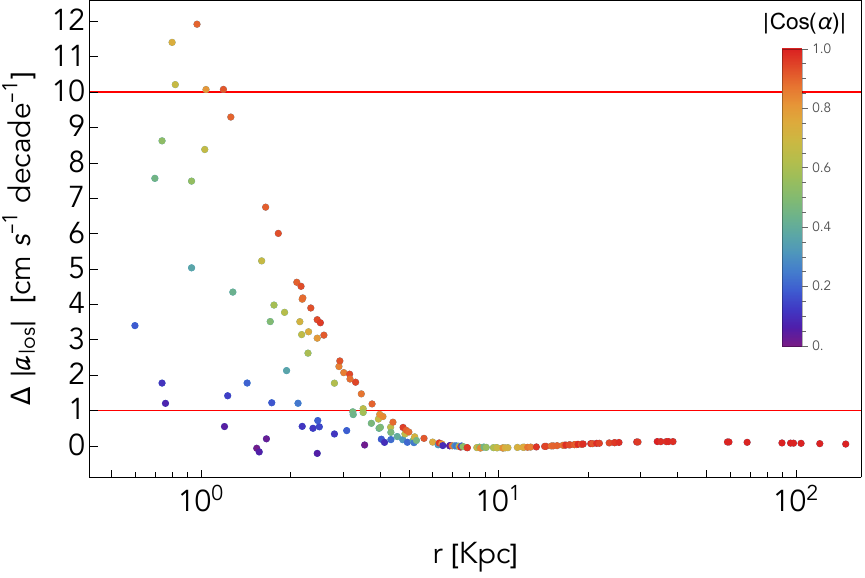} 
        \end{center}
        \caption{\textit{Left.} Total line of sight acceleration calculated with the Newtonian and Yukawa best fits Fig. \ref{RC_BestFit} for the adopted globular clusters. Red line denotes 10 cm $\text{s}^{-1} \text{decade}^{-1}$. \textit{Right.} Absolute difference between Newtonian and Yukawa gravity LOS acceleration for each globular cluster from the adopted dataset, assuming the best fits from Fig. \ref{RC_BestFit}. The colors denote the absolute value of $\cos(\alpha)$, Eq. \ref{cosalpha},  for each globular cluster. The GCs with $\Delta |a_{\rm los}| > 10 \, {\rm cm} \, {\rm s}^{-1} \, {\rm decade}^{-1}$ are Djor 2, Ter 4, Ter 6, NGC 6522 and Pal 6, see also Table \ref{tableIII}.  }
        \label{TotalAcce} 
\end{figure}

The LOS accelerations of the globular clusters in Ref.~\cite{2021MNRAS.505.5957B} sample versus galactocentric distance for the best fits of Fig.~\ref{RC_BestFit} are plotted in Fig.~\ref{TotalAcce}.
Given the large number of targets with high LOS acceleration, it is interesting to calculate the difference in LOS acceleration between Yukawa and Newtonian gravity. The targets with the highest difference could be candidates in a direct search of LOS accelerations for comparisons between Newtonian and Yukawa gravity, similarly to \cite{Quercellini:2008it, 2023JCAP...11..035F, 2024ApJ...974..223A}. We find that only five globular clusters have LOS acceleration differences higher than 10 cm $\text{s}^{-1} \text{decade}^{-1}$. These are shown in Table \ref{tableIII}  along with the differences in LOS acceleration and their distances to the Galactic center. From the remaining 160 globular clusters only 44 have LOS acceleration differences higher than 1 cm $\text{s}^{-1} \text{decade}^{-1}$. The complete distribution can be seen in Fig. \ref{TotalAcce}. Since we assume the same baryonic matter distribution for Newtonian and Yukawa gravity, it follows that the difference between LOS accelerations depends only the dark matter contribution
\begin{subequations}
    \begin{align}
    &\Delta | a_{\text{los}} |  = |\hat{\boldsymbol{u}} \cdot \nabla \Psi^N_{\text{dm}} (\boldsymbol{\theta}_1)| - |\hat{\boldsymbol{u}} \cdot \nabla \Psi^{MG}_{\text{dm}} (\boldsymbol{\theta}_2)|, \\
 &\Psi^{MG}_{\text{dm}} \equiv \Psi^N_{\text{dm}} + \Psi^Y_{\text{dm}},
\end{align}
\end{subequations}
where $\Psi^{N}_{\text{dm}}$,  $\Psi^Y_{\text{dm}}$ are defined in Eq. \ref{dmDef} and $\boldsymbol{\theta}_1, \boldsymbol{\theta}_2$ are the best fit points, Fig. \ref{RC_BestFit}. From the spherical symmetry of the dark matter contribution, it follows that 
\begin{subequations}\label{cosalpha}
    \begin{align}
    &\Delta | a_{\text{los}} |  = \cos (\alpha) \, \left|\frac{d \,\Psi^N_{\text{dm}} (\boldsymbol{\theta}_1)}{dr} \right| -\cos (\alpha) \left | \frac{d \, \Psi^{MG}_{\text{dm}} (\boldsymbol{\theta}_2)}{dr}\right |, \\
    &\cos(\alpha) \equiv \hat{\boldsymbol{u}} \cdot \boldsymbol{r}.
\end{align}
\end{subequations}
In summary, the difference is proportional to $\cos(\alpha)$ which is a projection effect. Figure \ref{TotalAcce} also illustrates the effect of $\cos(\alpha)$ for globular clusters. For an observer located at the Galactic center, the LOS coincides with the radial direction, and the figure would then display a single curve. Instead, we observe that the acceleration differences tend to vanish at large radii, albeit with some scatter at fixed values of $r$. This behavior arises from the projection onto the Solar system's LOS.
\begin{table}
    \centering
    \caption{Globular clusters (first 5 lines) and RR-Lyrae stars (last 5 lines) with the largest differences in LOS acceleration between Newtonian and Yukawa gravity. They are sorted with respect to the galactocentric radius. 
    The name of RR Lyrae stars refer to their unique source identifier from GAIA DR3 \cite{2023ApJ...944...88L}.}
    \renewcommand{\arraystretch}{1.5} 
    \begin{tabular}{ c c c} 
    \hline
    \hline
    Name & $\Delta |a_{\text{los}}| \, [\text{cm} \, \text{s}^{-1} \, \text{decade}^{-1}]$ & $r \, [\text{kpc}]$  \\ 
    \hline
    Djor 2 & $11.9$ &  $0.8$  \\ 
    Ter 4 & $10.2$ & 0.82 \\
    Ter 6 & 11.9 & 0.97 \\
    NGC 6522 & 10. & 1.04\\
    Pal 6 & 10. & 1.19 \\
    \hline
    4 056 428 080 895 480 704 & 12.8 & 0.37 \\
    4 057 433 996 626 472 576 & 12.4 & 0.38 \\
    4 060 264 380 043 171 072 & 12.3 & 0.43 \\
    4 057 350 502 400 434 176 & 12.4 & 0.48 \\
    4 060 254 106 473 320 704 & 12.7 & 0.50 \\
    \hline
    \end{tabular}
    \label{tableIII}
\end{table}

\begin{figure}
    \includegraphics[width=0.35\textwidth]{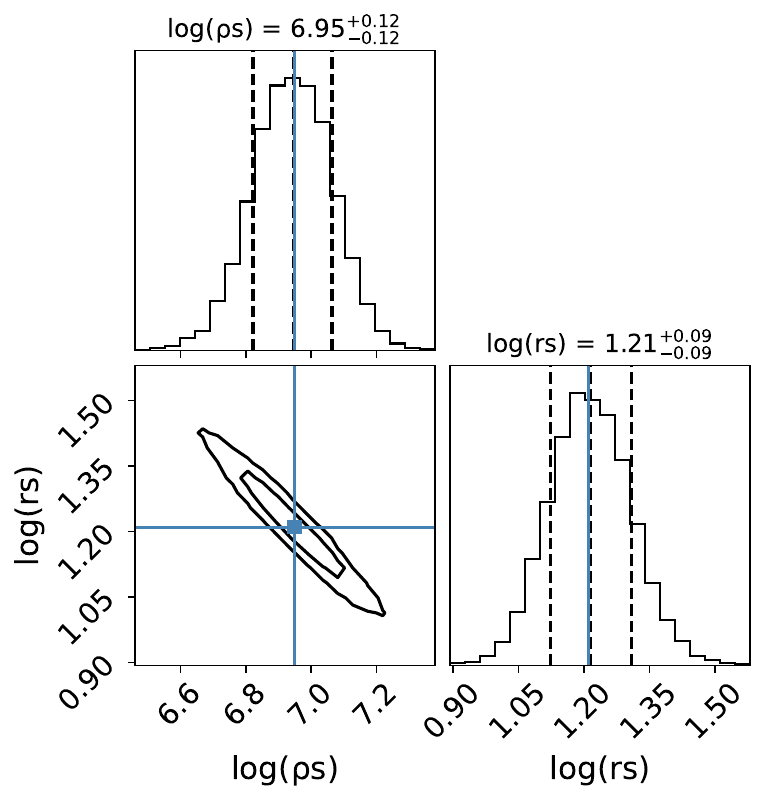}
    \includegraphics[width=0.64\textwidth]{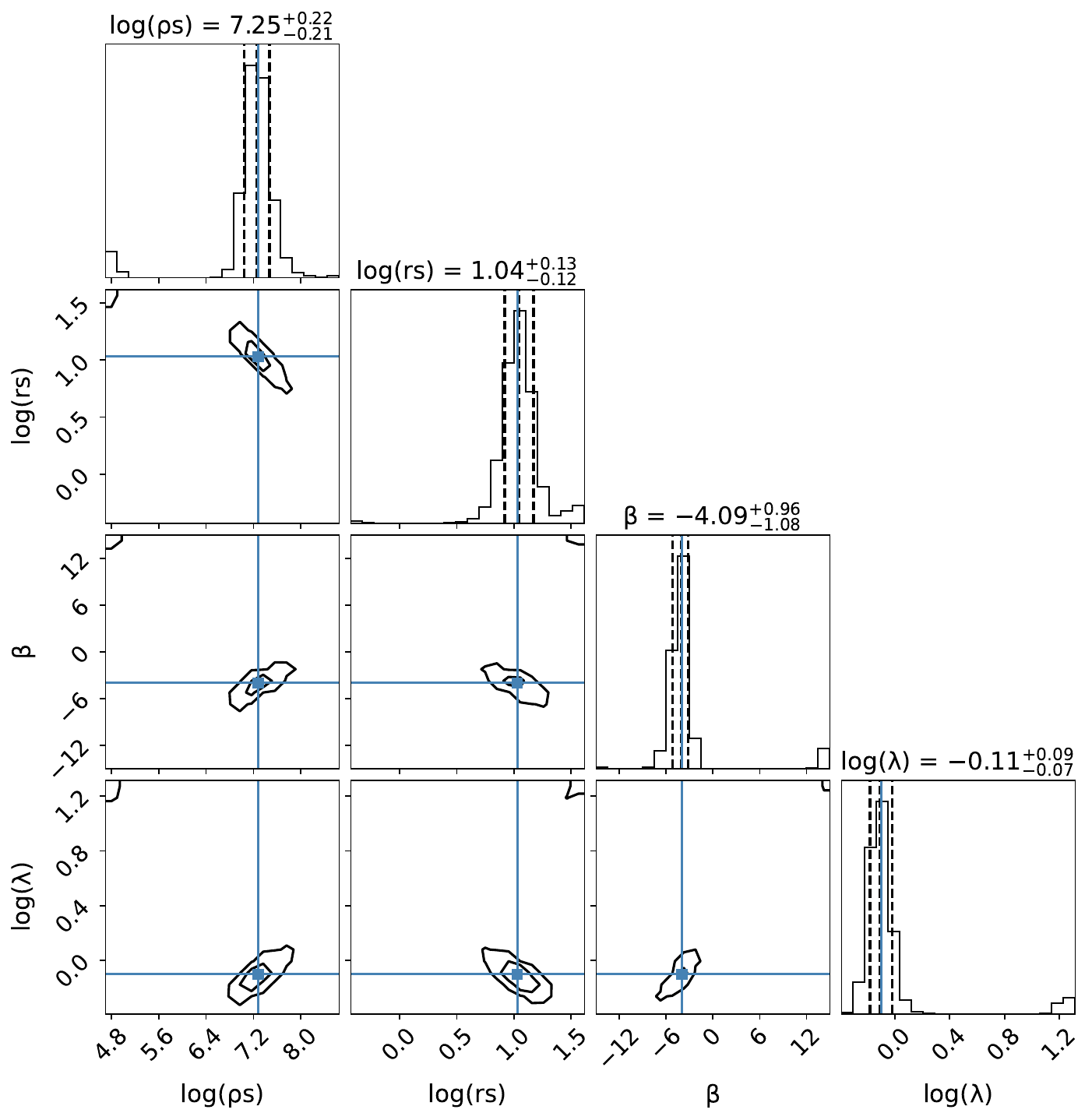}
    \caption{Newtonian (left) and Yukawa gravity (right) posteriors from globular clusters with precisions of 1 and $0.6\, \text{cm}\, \text{s}^{-1 }\, \text{decade}^{-1}$ respectively. Dashed lines denote $\{0.16, 0.5, 0.84\}$ quantiles and the blue lines denote the fiducial points as defined in Fig. \ref{RC_BestFit}.}
    \label{2plots}
\end{figure}

\begin{figure}
    \includegraphics[width=0.49\textwidth]{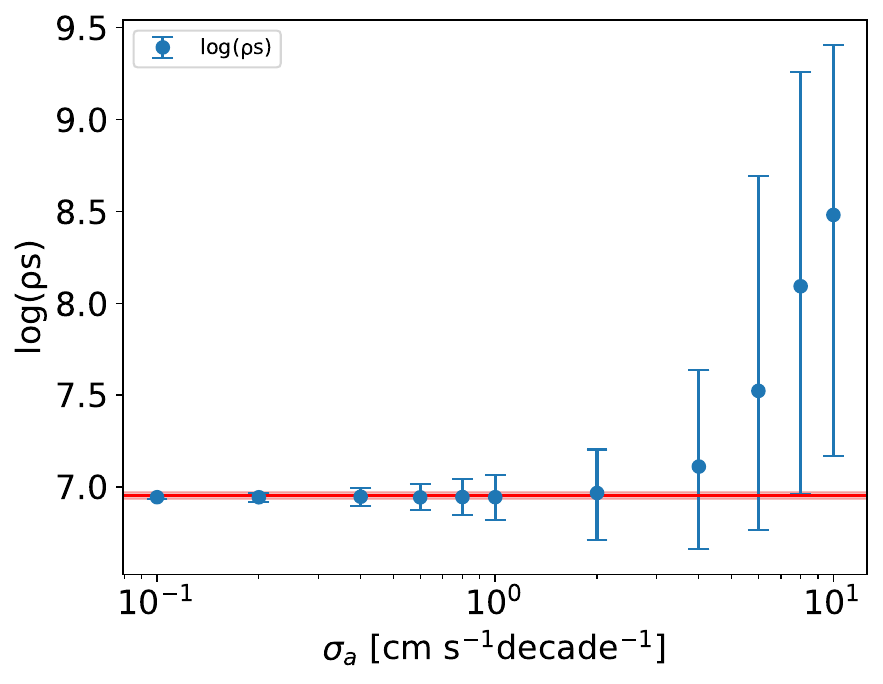} 
    \includegraphics[width=0.49\textwidth]{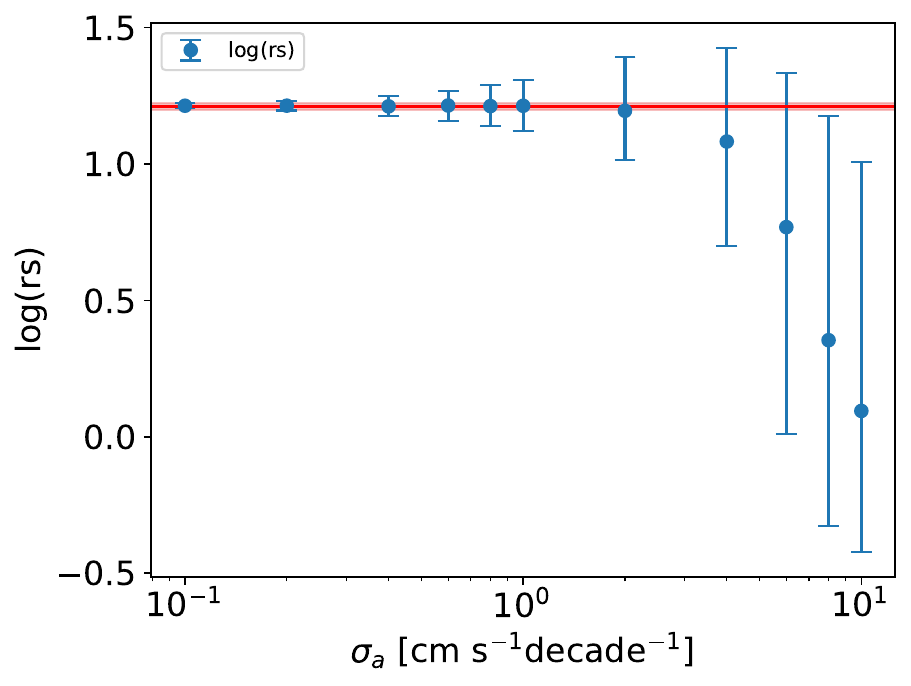}
    \caption{Marginalized errors from globular clusters for Newtonian gravity. The shaded red region denotes the $\{0.16, 0.84\}$ quantile interval from Galaxy rotation curve, the parameter errors correspond to the same quantile interval from globular clusters.}
    \label{UncertaintyGridNewton}
\end{figure}

\begin{figure}
    \begin{subfigure}[b]{0.5\textwidth} 
        \includegraphics[width=\textwidth]{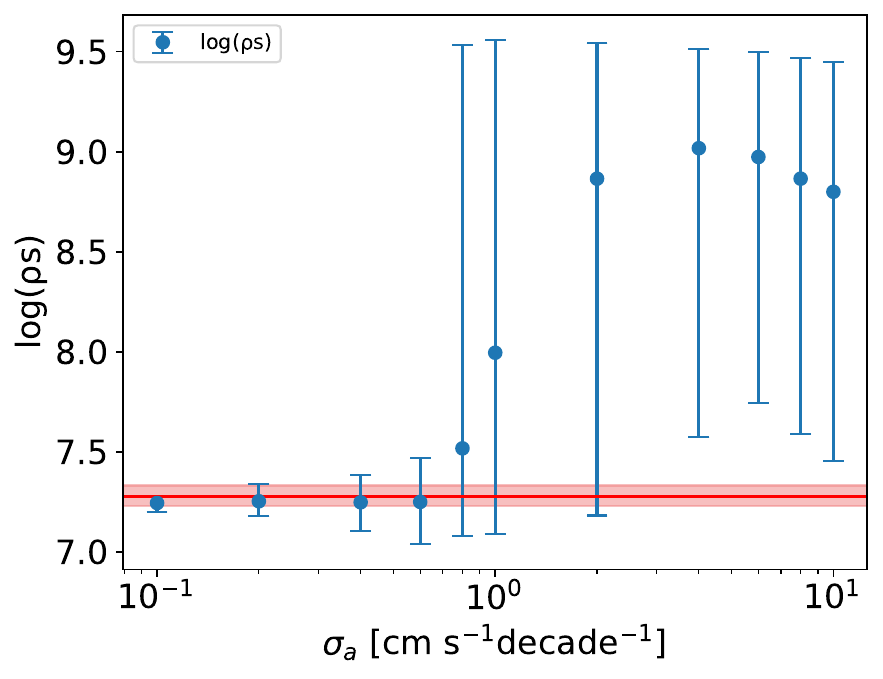}
    \end{subfigure}
    \hspace{1mm} 
    \begin{subfigure}[b]{0.5\textwidth} 
        \includegraphics[width=\textwidth]{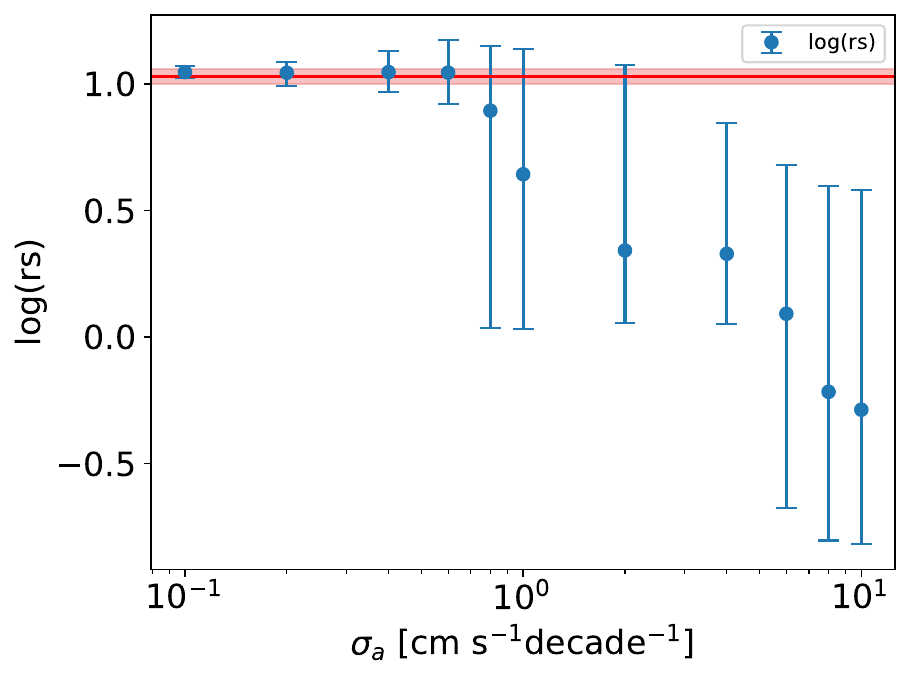}
    \end{subfigure}
    \\
    \begin{subfigure}[b]{0.5\textwidth} 
        \includegraphics[width=\textwidth]{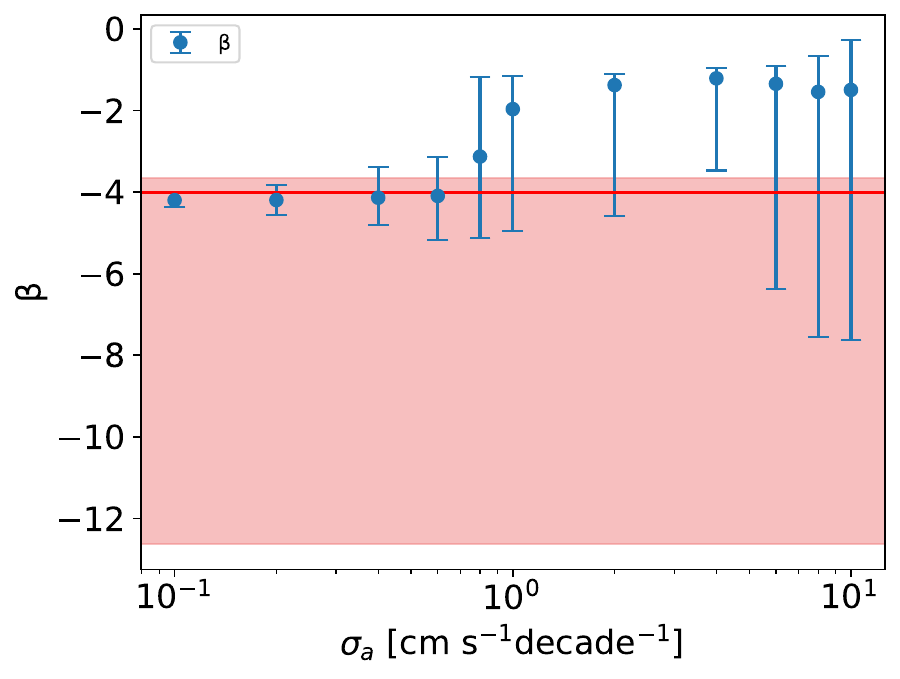}
    \end{subfigure}
    \hspace{1mm} 
    \begin{subfigure}[b]{0.5\textwidth} 
        \includegraphics[width=\textwidth]{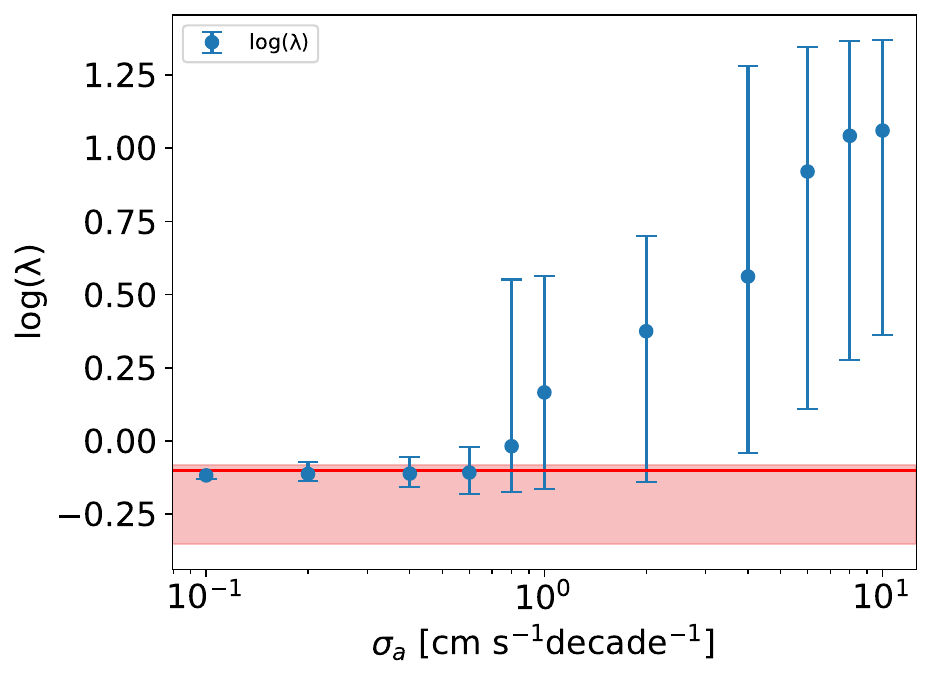}
    \end{subfigure}
    \caption{The blue data points with 1$\sigma$ errorbars denote the marginalized parameter results from globular clusters  data, for different $a_{\rm los}$ precision values ($\sigma_\text{a}$). The shaded red region denotes the 1$\sigma$ interval from the Galaxy rotation curve data and the red line the corresponding mean. Both the data points and the region refer to Yukawa gravity results.  The $1\sigma$ region is found from the 0.16 and 0.84 quantiles.}
    \label{UncertaintyGridYu}
\end{figure}

\begin{figure}
        \begin{center}
        \includegraphics[width=0.35 \textwidth]{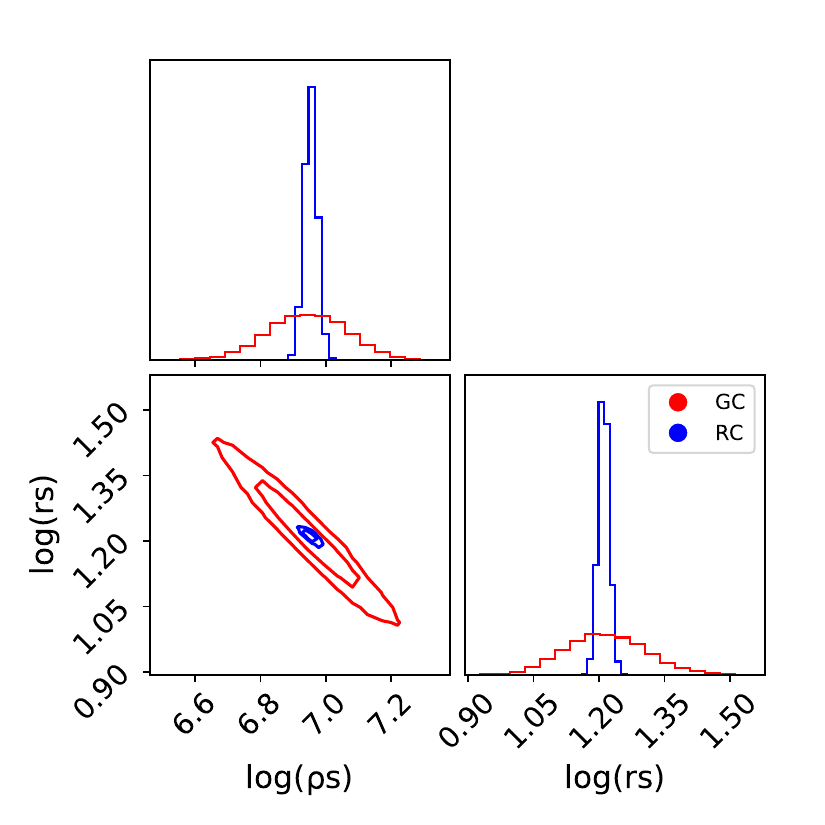}
        \includegraphics[width = 0.64 \textwidth]{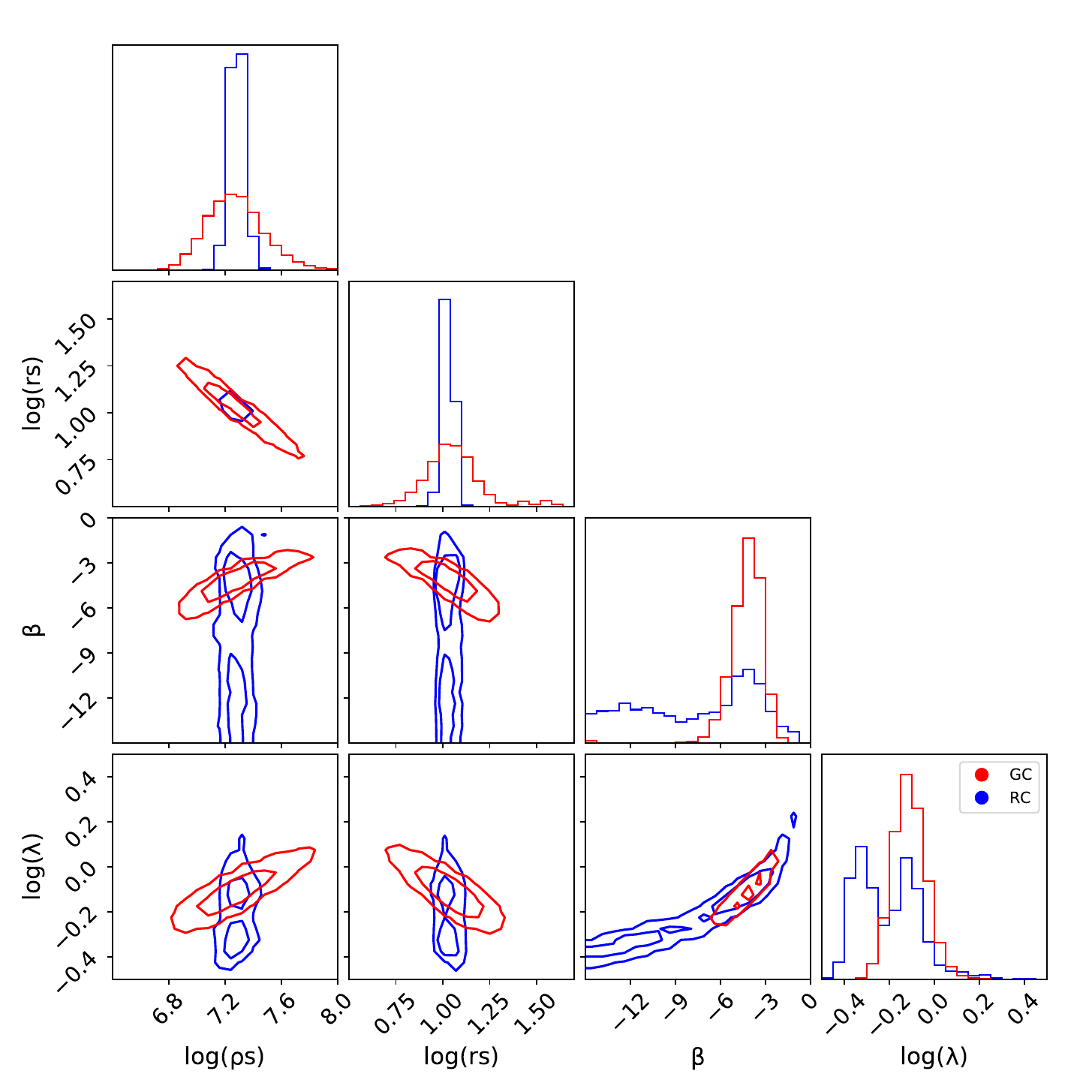}
        \end{center}
        \caption{ \textit{Left.} Newtonian Galaxy rotation curve (blue) and globular cluster LOS (red) posteriors with  $\sigma_a = 1 \, \text{cm}\, \text{s}^{-1 }\, \text{decade}^{-1}$. \textit{Right.} Yukawa Galaxy rotation curve (blue) and globular cluster LOS (red) posteriors with  $\sigma_a = 0.6 \, \text{cm}\, \text{s}^{-1 }\, \text{decade}^{-1}$.}
        \label{overplot1}
    \end{figure}

We started the Bayesian forecast assuming a precision of $\sigma_a  = 10\, \text{cm}\, \text{s}^{-1 }\, \text{decade}^{-1}$ and found it too large: the resulting posterior was non informative and prior dominated. We tested higher precisions, to see at which uncertainty the globular  cluster posteriors would be roughly as constraining as the rotation curve ones\footnote{We used $\sigma_a = (10, 8, 6, 4, 2, 1, 0.8, 0.6, 0.4, 0.2, 0.1) \,  \text{cm}\, \text{s}^{-1 }\, \text{decade}^{-1}$.}. For the Newtonian case at $\sigma_a \leq  1 \, \text{cm}\, \text{s}^{-1 }\, \text{decade}^{-1}$ the posterior no longer depends on the prior, whereas for Yukawa gravity $\sigma_a \leq 0.6 \, \text{cm}\, \text{s}^{-1 }\, \text{decade}^{-1}$  is necessary. The posteriors can be seen in Fig. \ref{2plots}. In order to obtain parameter constraints that are comparable to those from Galaxy rotation curve we found that one needs to set $\sigma_a = 0.4 \text{ cm}\, \text{s}^{-1 }\, \text{decade}^{-1}$ for Newtonian gravity. On the other hand, although the same precision is needed for NFW parameters in Yukawa gravity, our calculations show that $(\beta, \log \lambda)$ are already better constrained at $\sigma_a = 0.6 \text{ cm}\, \text{s}^{-1 }\, \text{decade}^{-1}$, as compared to the constraints from Galaxy rotation curve. These results are summarized in Figs. \ref{UncertaintyGridNewton} and \ref{UncertaintyGridYu}. We also provide in Fig. \ref{overplot1} a direct comparison between the Galaxy rotation curve and GC posterior at $\sigma_a = 0.6 \, \text{cm}\, \text{s}^{-1 }\, \text{decade}^{-1}$ for Yukawa gravity.  Figure \ref{overplot1} shows that the degeneracy directions in the LOS acceleration posterior for $(\beta, \log \rho_s),\, (\beta, \log r_s), \,(\log \lambda, \log \rho_s) \text{ and } (\log \lambda, \log r_s)$ differ from those obtained using rotation curve data alone. This suggests that a joint analysis combining rotation curves and LOS accelerations can provide tighter constraints on these parameters. Specifically, we find that for $\sigma_a = 1 \,  \text{cm}\,\text{s}^{-1}\, \text{decade}^{-1}$, the posterior is dominated by the prior in the parameters $(\log \rho_s, \log r_s)$, but not in $(\beta, \log \lambda)$. Therefore, even at this level of precision, a combined dataset can be informative, leveraging the constraining power of rotation curves for the NFW parameters and LOS accelerations for the Yukawa gravity parameters.

For completeness, we provide in table \ref{table_unique} the 1$\sigma$ confidence levels for the parameters of Newtonian and Yukawa gravity, obtained with the LOS accelerations of globular clusters for all values of $\sigma_a$ used in this work.

\subsection{RR Lyrae variable stars}\label{appendix}

In this section we investigate how the posteriors obtained from LOS accelerations benefit from an increased number of targets. We used RR Lyrae variable stars, found by the third Gaia data release \cite{2023ApJ...944...88L}. These stars have Galactocentric radius ranging from 0.2 to 138 kpc and the total number of stars is around $1.3 \times 10^5$.  Of course this sample includes both field stars, which are better tracers of the halo potential, and stars that belong to globular clusters, but in this preliminary assessment of feasibility we do not try to perform this distinction. We expect anyway that the RR-Lyrae stars associated to GCs are a subdominant population.

\begin{figure}
    \includegraphics[width=0.49\textwidth]{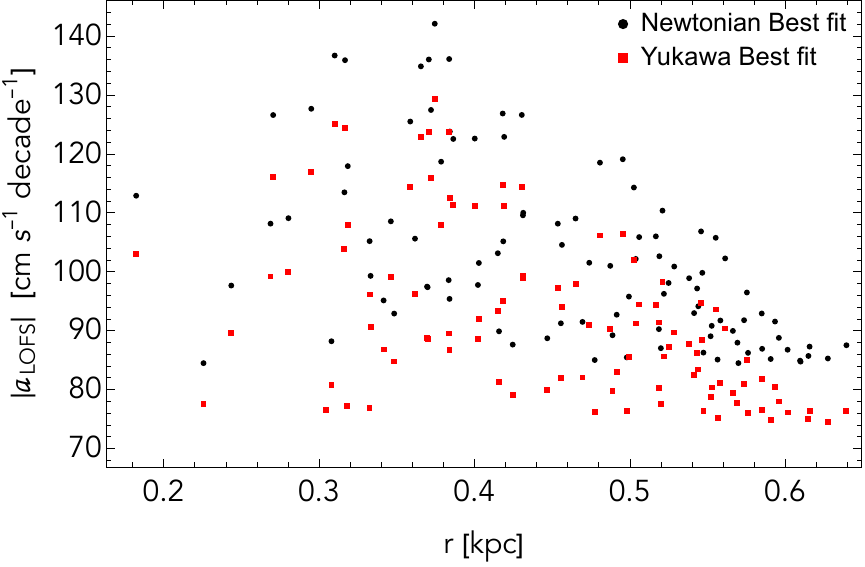}
    \includegraphics[width=0.49\textwidth]{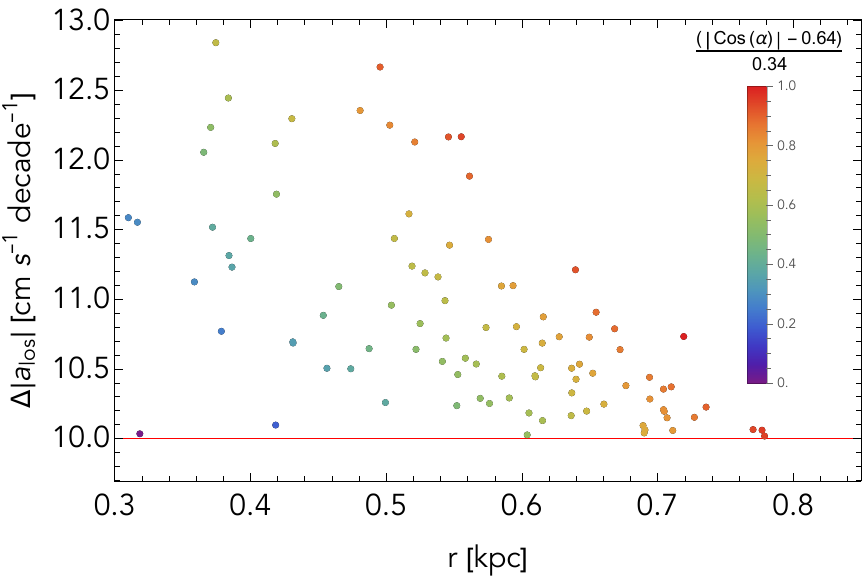}
    \caption{\textit{Left.} The $100$ RR-Lyrae stars with the largest LOS acceleration values. \textit{Right.} 100 highest differences in LOS acceleration between Newtonian and Yukawa gravity. The color variable, $0 \leq x \leq1$, is the normalized difference between the absolute value of $\cos( \alpha)$ and its minimum value. }
    \label{RRLYRAE100}
\end{figure}

Similarly to the last section, we calculated the total acceleration of these stars in Newtonian and Yukawa gravity. Our results show that the RR Lyrae stars with the highest accelerations have values around 140 $\text{cm}\, \text{s}^{-1} \, \text{decade}^{-1}$, bigger than the globular clusters with highest acceleration.
Moreover, the largest difference between Newtonian and Yukawa gravity is around 13 $\text{cm}\, \text{s}^{-1} \, \text{decade}^{-1}$ comparable to the largest differences in the globular cluster case. The difference between GCs and RR Lyrae stars arise because we have RR-Lyrae stars with $r \approx 0.2 \, \text{kpc}$, whereas the globular cluster sample has the smallest Galactic radius of $r \approx 0.8 $ kpc. We also note that there are 100 RR Lyrae stars with differences between the Newtonian and Yukawa gravity los acceleration that exceeds $10 \, \text{cm} \, \text{s}^{-1} \, \text{decade}^{-1}$, the first five of which are listed in Table \ref{tableIII}. These results are summarized in Fig. \ref{RRLYRAE100}.

The Bayesian analysis of the Newtonian case follows the Gaussian Likelihood in Eq. \ref{Likelihood}. However, given the computational cost of the analysis with more parameters we binned the data for the Yukawa gravity analysis. 
We employed a logarithmic binning of the radial coordinate $r$, spanning the range $[0.2, 138]$ kpc, using multiplicative steps of 1.1. For $\cos(\alpha)$ we used linear bins of length 0.04, where
\begin{align}
    \cos(\alpha) \equiv \boldsymbol{\hat{r}} \cdot \boldsymbol{\hat{u}}.
\end{align}

\begin{figure}
    \includegraphics[width=0.35\textwidth]{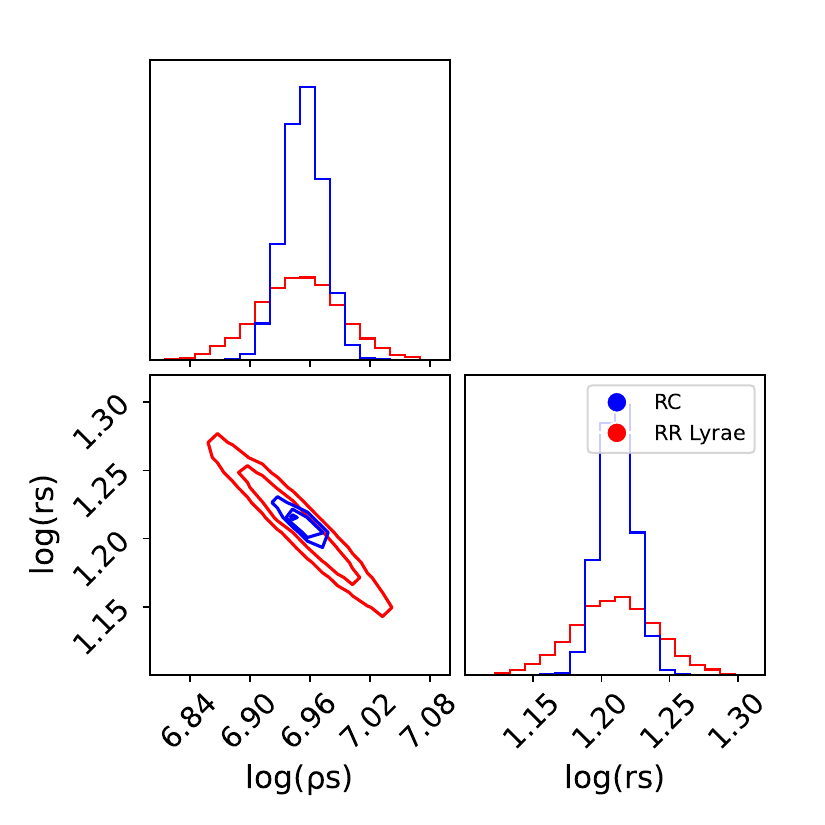}
    \includegraphics[width=0.64\textwidth]{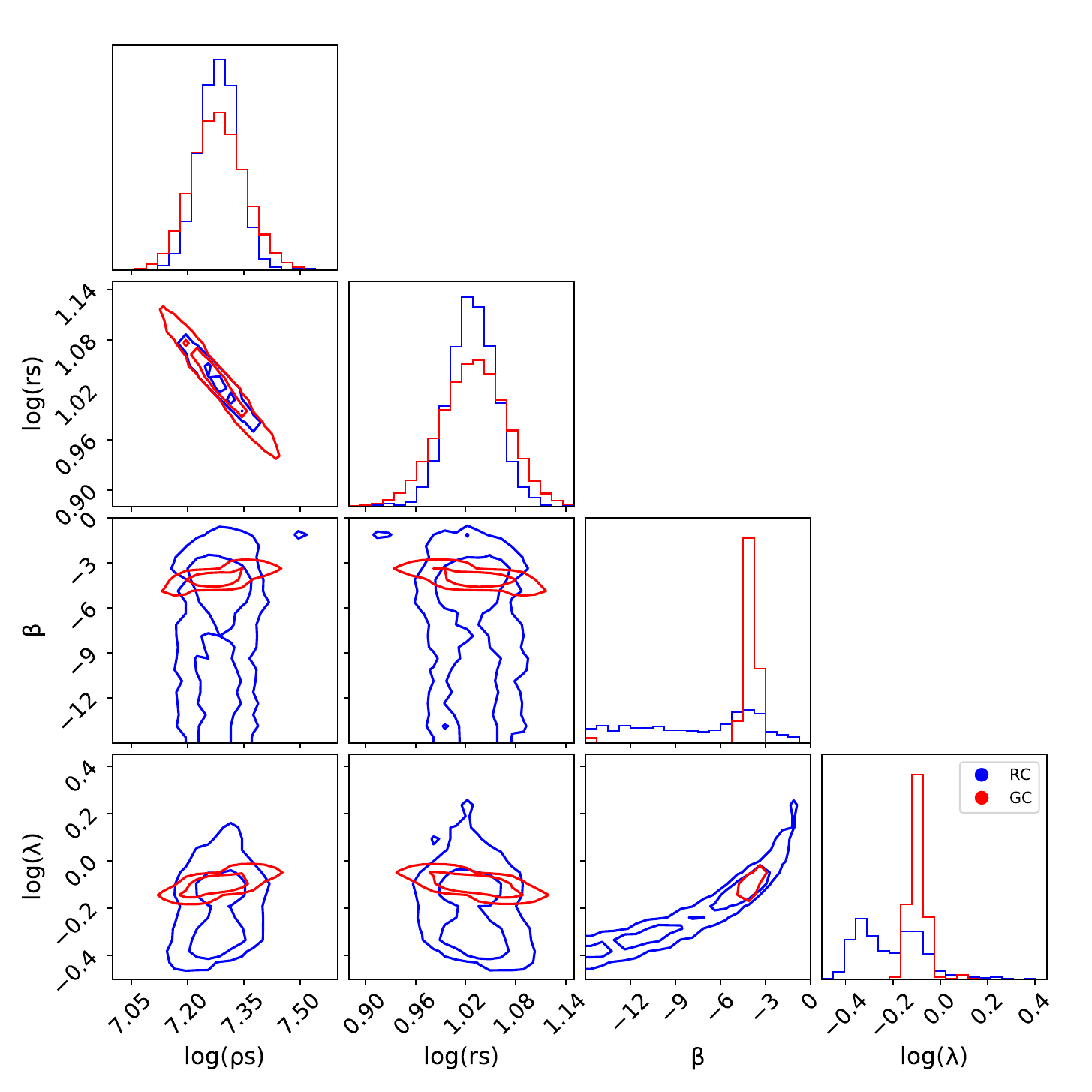}
    \caption{\textit{Left}. Newtonian Galaxy rotation curve (blue) and RR-Lyrae LOS (red) posteriors with  $\sigma_a = 10 \, \text{cm}\, \text{s}^{-1 }\, \text{decade}^{-1}$. \textit{Right}. Yukawa gravity Galaxy rotation curve (blue) and RR-Lyrae LOS posteriors with $\sigma_a = 10 \, \text{cm}\, \text{s}^{-1 }\, \text{decade}^{-1}$.}
    \label{overplot_RRLyrae}
\end{figure}

Our calculations show that the considerable number of stars,  $1.36 \times 10^5$, is enough to provide valid constraints even at $\sigma_a = 10 \, \text{cm}\, \text{s}^{-1 }\, \text{decade}^{-1}$, which is not possible with globular clusters. The posteriors are not prior dominated and their constraints, although not as good as rotation curve ones for NFW parameters, are comparable. In Table \ref{table_unique} we show the comparison of $1\sigma$  confidence levels for both gravitational theories using rotation curves and RR-Lyrae stars. In particular we found that for Yukawa gravity, although the constraints on the NFW parameters are not as good as the rotation curve ones, the constraint in Yukawa gravity parameters are far better than their counterpart from Galaxy rotation curve, as can be seen in Table \ref{table_unique}. Focusing on the Yukawa parameters $\beta,\lambda$, we forecast roughly a 10\% and a 30\% constraint, respectively:
\begin{equation}
    \beta= -3.98^{+0.33}_{-0.37}\,,\quad \lambda=-0.1^{+0.03}_{-0.03}.
\end{equation}
Since the coupling can be distinguished from zero to a high level of significance, we expect that even a higher $\sigma_a$ can achieve a discrimination between standard gravity and Yukawa gravity.
These results are also summarized in Fig. \ref{overplot_RRLyrae} where we plot the posteriors from Galaxy rotation curve and LOS accelerations of RR-Lyrae stars. 

\section{Conclusions}

We have studied the impact of two complementary kinds of targets for parameter estimation with LOS accelerations. 
Similarly to \cite{Quercellini:2008it}, we started by considering globular clusters as targets. Among the advantages for globular clusters as targets, we note the possibility of using many stars to infer the average motion of the cluster, improving the precision compared to single-star measurements. In addition to that, we have also used RR Lyrae stars, which are abundant in Milky Way's halo. These represent complementary kinds of targets, and the extent to which they will be useful in the future  depends on the experimental feasibility of high precision spectrography as applied for theses targets.

Our results show that $10 \,  \text{cm}\, \text{s}^{-1 }\, \text{decade}^{-1}$ is not enough precision to produce posteriors as competitive as the ones from the Galaxy rotation curve, if one considers only globular clusters. According to our analysis, a precision of at least $0.6 \,  \text{cm}\, \text{s}^{-1 }\, \text{decade}^{-1}$ would be necessary to get something comparable. On the other hand, using RR Lyrae stars  we can show that a precision of $10 \,  \text{cm}\, \text{s}^{-1 }\, \text{decade}^{-1}$ is enough to  recover constraints  on the Yukawa and halo parameters comparable to those  from Galaxy rotation curve analysis. It is worth noting that our results for RR Lyrae stars could be further improved by using the larger catalog \cite{Clementini_2023}, which contains approximately twice the number of stars we considered here.

In the context of direct measurements of different LOS accelerations, we remark the importance of selecting targets near the Galactic center. As evidenced by globular clusters and RR Lyrae stars, the difference between Newtonian and Yukawa gravity goes to zero far away from the Galactic center, a consequence of the Yukawa correction being exponentially decaying. Therefore in a direct search for differences it is crucial to select targets closer to the center.
         
\bigskip
\noindent
\textbf{Acknowledgements} \\
FASB thanks the Institute for Theoretical Physics at Heidelberg University for hospitality, where a large part of this work was developed. FASB acknowledges FAPES and CAPES for support. 
LA acknowledges support by the Deutsche Forschungsgemeinschaft (DFG, German Research Foundation) under Germany's Excellence Strategy EXC 2181/1 - 390900948 (the Heidelberg STRUCTURES Excellence Cluster) and under Project  554679582 "GeoGrav: Cosmological Geometry and Gravity with non-linear physics". 
DCR thanks \textit{Centro Brasileiro de Pesquisas Físicas} (CBPF) and \textit{Núcleo de Informação C\&T e Biblioteca} (NIB/CBPF) for hospitality, where part of this work was done. He also acknowledges CNPq (Brazil), FAPES (Brazil) and \textit{Fundação de Apoio ao Desenvolvimento da Computação Científica} (FACC, Brazil) for partial support.
The authors acknowledge the use of the computational resources provided by the Sci-Com Lab of the Department of Physics at UFES, which is funded by FAPES and CNPq.

\printbibliography
\end{document}